\documentstyle[12pt,aasms4]{article}
\tighten

\def\lsim{\lower.5ex\hbox{$\; \buildrel < \over \sim \;$}}
\def\gsim{\lower.5ex\hbox{$\; \buildrel > \over \sim \;$}}
\def\lax    {\ifmmode{_<\atop^{\sim}}\else{${_<\atop^{\sim}}$}\fi}
\def\gax    {\ifmmode{_>\atop^{\sim}}\else{${_>\atop^{\sim}}$}\fi}
\def\etal{{\it et al.\/} }
\def\gtorder{\mathrel{\raise.3ex\hbox{$>$}\mkern-14mu
             \lower0.6ex\hbox{$\sim$}}}
\def\ltorder{\mathrel{\raise.3ex\hbox{$<$}\mkern-14mu
             \lower0.6ex\hbox{$\sim$}}}

\def\pmb#1{\setbox0=\hbox{#1}%
  \kern-0.015em\copy0\kern-\wd0
  \kern0.03em\copy0\kern-\wd0
  \kern-0.015em\raise0.0433em\box0 }

\begin{document}

\title{ Interpretation of $\sim 35$ Hz QPO in the Atoll Source 4U
1702-42 as a Low Branch of the Keplerian Oscillations Under the Influence
of the Coriolis Force}

\author{Vladimir Osherovich}
\affil{NASA/Goddard Space Flight Center/RITSS, Greenbelt MD 20771 USA;
vladimir@urap.gsfc.nasa.gov}

\author{Lev Titarchuk}
\affil{NASA/ Goddard Space
Flight Center, Greenbelt MD 20771, and George Mason University/CSI, USA;
titarchuk@lheavx.gsfc.nasa.gov}

\vskip 0.5 truecm


\begin{abstract}

The recent model of quasi-periodic oscillations in neutron star binaries
(Osherovich and Titarchuk 1999, Titarchuk and Osherovich 1999) has
suggested the existence of two branches of QPOs due to the influence of
Coriolis force on the linear Keplerian oscillator: one branch with frequencies
$\sim 400-1200$ Hz and another branch with frequencies an order of magnitude
lower. The frequencies of the high branch $\nu_h$ hold a hybrid frequency
relation with the Keplerian frequency $\nu_K$:
${\nu_h}^2={\nu_K}^2+ {\Omega}^2/{\pi}^2$, where $\Omega$ is the
rotational frequency of the star's magnetosphere.
The frequency of the low branch is $\nu_L=(\Omega/\pi)(\nu_K/\nu_h)
\sin\delta$, where $\delta$ is a small angle between $\bf{\Omega}$ and
the vector normal to the plane of Keplerian oscillations.
The observations of the source 4U 1702-42 (Markwardt et al 1999) have shown
that the centroid of the $\sim 35$ Hz QPO tracks the frequency of
the kilohertz oscillations. We interpret the $\sim 35$ Hz
oscillations as $\nu_L$ and find $\delta=3.9^o \pm 0.2^o$. Our results
make 4U 1702-42 the second source (after Sco X-1) for which the theoretically
derived lower branch is identified (within our model) and $\delta$ is
calculated. The inferred angle $\delta$  stays approximately the same over
the significant range of $\nu_K$ (650 - 900 Hz), as expected from the model.
Based on our model we present a classification of QPO frequencies in
the source 4U 1702-42 observed above and below $\nu_L$.

\end{abstract}


\section{Introduction}

The discovery of kilohertz quasiperiodic oscillations (QPOs) in the low
mass X-ray neutron star binaries (Strohmayer \etal 1996;
Van der Klis \etal 1996) was followed by
similar results for nineteen sources (Van der Klis \etal 1998, Strohmayer,
Swank and Zhang 1998).
For most of them, the Rossi X-Ray Timing Explorer (RXTE) mission showed
the persistence of twin peaks in the spectrum ($\sim400-1200$ Hz). In the
lower part of the spectrum of Sco X-1, van der Klis \etal (1997) found two
branches $\sim45$ and $90$ Hz which correlate with the frequencies of the
above-mentioned twin peaks. The
nature of the twin peaks has been discussed intensively. The apparent
constancy of the difference $\Delta\nu$ for the twin peaks in some sources led
to the beat-frequency interpretation
(a concept originated by Alpar and Shaham 1985). Below,
we describe difficulties  of this model following the recent assessment by
Mendez and van der Klis (1999). Within the beat-frequency model,
the higher peak in the kHz range is
identified with the Keplerian frequency (van der Klis \etal 1996). The
lower twin peak in this model occurs as a result of beating between the high
peak and the spin frequency of the neutron star $\nu_{spin}$ which is
believed to be observed during type 1 X-ray bursts as $\nu_{burst}$
(or half that) (Strohmayer \etal 1996, Miller,
Lamb \& Psaltis 1998). Within the original beat-frequency model,
$\Delta\nu$ and $\nu_{spin}$ should remain constant.
The observed $20-30\%$ variation of $\Delta\nu$
(Van der Klis \etal 1997, Mendez \etal 1998, Ford \etal 1998, Mendez \&
van der Klis 1999) undermined the beat-frequency interpretation.
Convincingly, Mendez and van der Klis (1999) showed that for atoll source
4U 1728-34, $\Delta\nu$ is not equal to
$\nu_{burst}$ even for the lowest inferred mass accretion rate. In their
words, this  seems to rule out the simple beat-frequency interpretation of
the kHz QPOs in LMXBs and some of the modifications introduced to explain the
results of a variable $\Delta\nu$.

A completely different paradigm has been proposed by Osherovich and
Titarchuk (1999) and by Titarchuk and Osherovich (1999)
(hereafter OT99 and TO99, respectively) to explain QPOs in neutron stars.
The new model is based on the idea that the twin
peaks occur as a result of the Coriolis force imposed on a
one-dimensional Keplerian oscillator. The second section of our Letter
contains the assumptions and predictions of the model.
Verification of the predictions concerning the low frequency branch for the
source 4U 1702-42 is presented in section 3.
Discussion and summary follow in the last section.

\section{Oscillations in the Disk and the Magnetosphere Surrounding the
Neutron Star}

The following assumptions constitute our model (OT99 and TO99):

a) The lower frequency of the twin
kHz QPO is the Keplerian frequency
\begin{equation}
\nu_K={1\over{2\pi}}\left({GM}\over {R^3}\right)^{1/2}
\end{equation}
where G is the gravitational constant, M is the mass of the neutron star
and R is the radius of the corresponding Keplerian orbit.

b) There is a boundary layer:
transition region between the NS surface and the first Keplerian orbit.
Within this layer, the radial viscous oscillations $\nu_{\rm v}$ are maintained.
There is also a diffusive propagation process of the perturbations
generated in this layer and  characterized by a break frequency
$\nu_b$ related to $\nu_{\rm v}$. Outside of this layer, the disk has Keplerian
radial oscillations with $\nu_K$ (see also Titarchuk, Lapidus \& Muslimov
1998).

 c) The magnetosphere which surrounds the neutron star has a rotational
frequency $\bf\Omega$ which is not perpendicular to the disk.

d) inhomogeneities (hot blobs) thrown into the vicinity of the disk
participate in the radial Keplerian oscillations but simultaneously are
subjected to the Coriolis force associated with the differential rotation of
the magnetosphere.

The consequences of the model are straightforward. The problem of a
linear oscillator in the frame of reference rotating with rotational frequency
$\bf\Omega$ is known to have an exact solution describing two branches of
oscillations
\begin{equation}
\nu_h^2=\nu_K^2+\left(\Omega\over\pi\right)^2
\end{equation}
\begin{equation}
\nu_L=(\Omega/\pi)(\nu_K/\nu_h)\sin\delta
\end{equation}
where $\delta$ is the angle between $\bf\Omega$ and the plane of the
Keplerian oscillations. Formulas (2) and (3) are derived under the
assumption of
small $\delta$ (OT99 and references therein). Thus, the high
frequency of the observed twin kHz peaks has been interpreted as the
upper hybrid frequency $\nu_h$ or the frequency of the high branch of the
Keplerian oscillator under the influence of the Coriolis force associated
with magnetospheric rotation.
The existence of the lower branch ($\nu_L$) is a prediction of the
model.
Oscillations $\sim 45$ Hz and 90 Hz in Sco X-1 reported by van der Klis
\etal (1997) are found to fit formula (3) for $\delta=5.5 \pm0.5^o$ when
interpreted as 1st and 2nd harmonics of $\nu_L$ (OT99, TO99).
In the low part of QPO spectra according to TO99 there are two more 
frequencies (namely the break frequency $\nu_b$ and frequency of viscous 
oscillations $\nu_{\rm v}$) related to each other
\begin{equation}
\nu_b=0.041\nu_{\rm v}^{1.61}.
\end{equation}
We now consider the application of our model  to source 4U 1702-42 studied in
detail by Markwardt, Strohmayer \& Swank (1999).

\section{Determination of $\delta$ for the Source 4U 1702-42}

For Sco X-1, we have shown that the solid body rotation
$\Omega=\Omega_0=const$ is a reasonable first order approximation (OT99).
Theoretically, $\Omega$ depends on the magnetic structure of the neutron
star's magnetosphere. Within the second order approximation, the $\Omega$
profile has a slow variation as a function of $\nu_K$, which we modeled for
Sco X-1 and for the source 4U 1608-52 in OT99. For the source 4U 1702-42, there
are not enough simultaneous measurements of $\nu_h$ and $\nu_K$ to
reconstruct the $\Omega$ profile. Thus, we restrict ourselves to the first
order approximation and adopt $\Omega/{2\pi}=\Omega_0/{2\pi}=380 \pm 7$ Hz,
found in  OT99. The observed $\nu_K$
and $\nu_h$ used in OT99 for the calculation of $\Omega_0/{2\pi}$
according to formula (2) are presented in Table 1 which contains the data
kindly provided by C. Markwardt.
Then from formulas (2) and (3) we have
the expression for
$\delta$, namely
\begin{equation}
\delta=\arcsin\left[(\pi/\Omega_0)
(\nu_K^2+\Omega_0^2/{\pi}^2)^{1/2}(\nu_L/\nu_K)\right]
\end{equation}
where $\nu_K$ is the observed Keplerian frequency (the second from the
top kHz QPO) and $\nu_L$ is the alleged frequency of the low Keplerian
branch.
Within the first order approximation, the angle $\delta$ should stay the
same for all $\nu_K$, since $\delta$ is effectively the angle between the
equatorial plane of the disk and the plane of the magnetic equator.
The first two columns of Table 2 present frequencies observed by
Markwardt \etal
(1999) for the source 4U 1702-42 with the corresponding date of the
measurements.
These two columns repeat the first two columns of the table in Markwardt
\etal (1999) with the addition of data for July 30, 1997 taken from the text
of the same paper.
In the third column of our Table 2, we give our theoretical
interpretation for the observed frequency peaks: K stands for the Keplerian
frequency, L for the frequency of the lower branch of the Keplerian frequency,
b for the break frequency. According to TO99,  the relation between 
  the  frequency of viscous oscillations 
$\nu_{\rm v}$ (symbol v in our table) and  frequency $\nu_b$ (Eq. 4) 
including equation (5) will serve  as a tool for the identification of 
$\nu_b$ and $\nu_{\rm v}$.
The last column of Table 2 contains $\nu_h$ calculated according to
formula (2) with $\Omega_0/{2\pi}=380$ Hz and also the angle $\delta$
(calculated according to formula (5) for those cases when 
$\nu_K$ and $\nu_L$ are measured simultaneously).
The resulting values of $\delta$ are shown in Figure 1.
 Indeed, the angle
$\delta$ shows little variation with $\nu_K$ and as expected is rather
small
\begin{equation}\
\delta=3.9^o \pm 0.2^o
\end{equation}
With knowledge of $\delta$ for the source 4U 1702-42, we attempted to
interpret the remaining observed frequencies in Table 2.
For July 26.60-26.93, $\nu_K$ and $\nu_h$
are not present. Three frequencies 10.8, 32.5 and 80.1 Hz were observed
on this day. 
If 32.5 Hz should be the $\nu_L$ then the second harmonic $2\nu_L=65$ Hz. But
instead we have 80.1 Hz, which is significantly higher than the expected
$2\nu_L$. On the other hand, from the classification Figure 4 presented in
TO99, we know that in the vicinity of $\nu_K=800-900$ Hz, the two
frequencies $\nu_L$ and $\nu_{\rm v}$ may come close to one another.
Thus assuming that $\nu_{\rm v}=32.5$
Hz, we find that $\nu_b=10.8$ Hz satisfies the theoretical relation (4)
within the observational errors. We identify $2\nu_L=80.1$ Hz and also
$2\nu_L=85.5$ Hz for July 21 and $\nu_b=12$ Hz for the same day.
From formula (3)  we derive $\nu_K$ for both
cases assuming the fixed $\delta$ and
$\Omega/2\pi=380$ Hz. This identification holds best for $\delta=4.3^o\pm
0.1^o$ which is the angle derived for $\nu_K=769$ Hz and $\nu_L=40.1$ Hz on
July 30 (see the last line of Table 2). The consistency of our
identification is shown in Figure  2 where $\nu_h$, $\nu_L$, $2\nu_L$ and 
$\nu_b$ are plotted as functions of $\nu_K$ (solid lines are theoretical).
Figure 2 shows that
as expected
$\Delta\nu=\nu_h - \nu_K$ decreases as $\nu_K$ increases. With the small
number of observational points, the theoretical curves in  Figure 2 are, in 
fact, a prediction of our model which can be viewed as a challenge for those 
observers who choose to extend the  observational base for the source 4U 
1702-42 to verify our model. It is possible that the difference between $\delta=3.8^o$ and observed
$\delta=4.3^o$ is attributed to $\Omega=const-$assumption. Further measurement
of $\nu_h$ and $\nu_K$  shall allow to reconstruct the $\Omega-$profile and
improve our predicative capability.

\section{Discussion and Conclusions}

Our model (OT99 and TO99) reveals  the physical nature of $\sim 35$ Hz
oscillations in the source 4U 1702-42. As a low Keplerian branch of
oscillations in the rotating frame of reference, this phenomenon should have an
observational invariant - namely the angle $\delta$. Within the first order
approximation this angle can be viewed as a global parameter describing the
inclination of the magnetospheric equator to the equatorial plane of the disk.
Measured locally for different radial distances
(therefore different $\nu_K$), $\delta$ may vary considerably unless indeed
the observed oscillations correspond to the predicted low
Keplerian branch. The constancy of $\delta$ shown in Figure 1 allows us
to interpret the $\sim 35$ Hz oscillations as the low branch described by
equation (3).
Knowledge of the angle $\delta$ is essential for the classification of
the QPO resonances and the understanding of the physical nature of this
phenomenon. The parameter $\delta$ is critical in evaluating
the differences between the spectra of
different sources. For Sco X-1, oscillations with frequency
$\sim 45$ Hz are identified as belonging to the  $\nu_L$ branch (OT99).
The higher observed frequencies of the $\nu_L$ for Sco X-1  are mainly
a result of a larger angle $\delta=5.5^o \pm 0.5^o$.

We believe that the angle $\delta$ as a fundamental geometric parameter
of the neutron star magnetosphere will be found eventually for other sources
with kHz QPO. The source 4U 1702-42 is the second source for which $\delta$ has
been inferred.

The authors are grateful to J. Fainberg and R.G. Stone for discussions
and suggestions. The comments of the referee which led to an essential
improvement of the paper are appreciated.

\clearpage
 \begin{table}
\centering
\begin{tabular}{|ccc|}
\hline
{$\nu_K$ (Hz)} & $\nu_h$ (Hz)& $\Omega/{2\pi}$ (Hz)\\
\hline
657. & $1000.\pm 7$ & $377.\pm5$\\
702. & $1038.\pm7$  & $382.\pm5$\\
769. & $1085.\pm 11$ & $383.\pm8$\\
\hline
\end{tabular}
\caption{Observed twin peak frequencies ($\nu_K$ and $\nu_h$) and
calculated rotational frequency $\Omega/{2\pi}$ for 4U 1702-42
(courtesy C. Markwardt).}
\end{table}
\clearpage

 \begin{table}
\centering
\begin{tabular}{|cccc|}
\hline
{Day of July, 1997} & Frequency (Hz) & Classification & Calculation \\
\hline
19.37-19.86 & 902. & K & $\nu_h=1179$ Hz\\
            & $39.\pm3.4$ & L & $\delta=3.8^o$\\
\hline
21.02-21.17 & $85.6\pm3.7$ & 2L & \\
            & $12.\pm1$ & b & \\
\hline
26.60-26.93 & $80.1\pm0.5$ & 2L &\\
            & $32.5\pm1.2$ & v & \\
            & $10.8\pm0.4$ & b & \\
\hline
30.46-30.72 & 722. & K & $\nu_h=1048$ Hz\\
            & $\Delta\nu=333.\pm5$ & $\nu_h-\nu_K$ & $\delta=3.8^o$\\
            & $34.7\pm 0.6$  & L & \\
\hline
30 & 657. & K & $\nu_h=1005$ Hz\\
    & $32.9\pm0.6$ & L & $\delta=3.8^o$\\
\hline
30 & 702. & K & $\nu_h=1035$ Hz\\
    & $34.8\pm0.6$ & L & $\delta=3.8^o$\\
\hline
30 & 769. & K & $\nu_h=1081$ Hz\\
   & $40.1\pm0.8$ & L & $\delta=4.3^o$\\
\hline
\end{tabular}
\caption{Observed frequencies of 4U 1702-42 and possible theoretical
interpretation.  Angle $\delta$ is calculated from formula (5). 
Observations are cited from Markwardt et al. 1999}.
\end{table}

\clearpage
\begin{figure}
\caption{ Inferred angle $\delta$ between the rotational frequency
$\bf\Omega$ and
the normal to the plane of Keplerian oscillations as a function of
observed
Keplerian frequency for the source 4U 1702-42.
\label{Fig.1}}
\end{figure}

\begin{figure}
\caption{ Classification of QPO in the atoll source 4U 1702-42. Solid
lines are theoretical curves.
\label{Fig.2}}
\end{figure}

\noindent
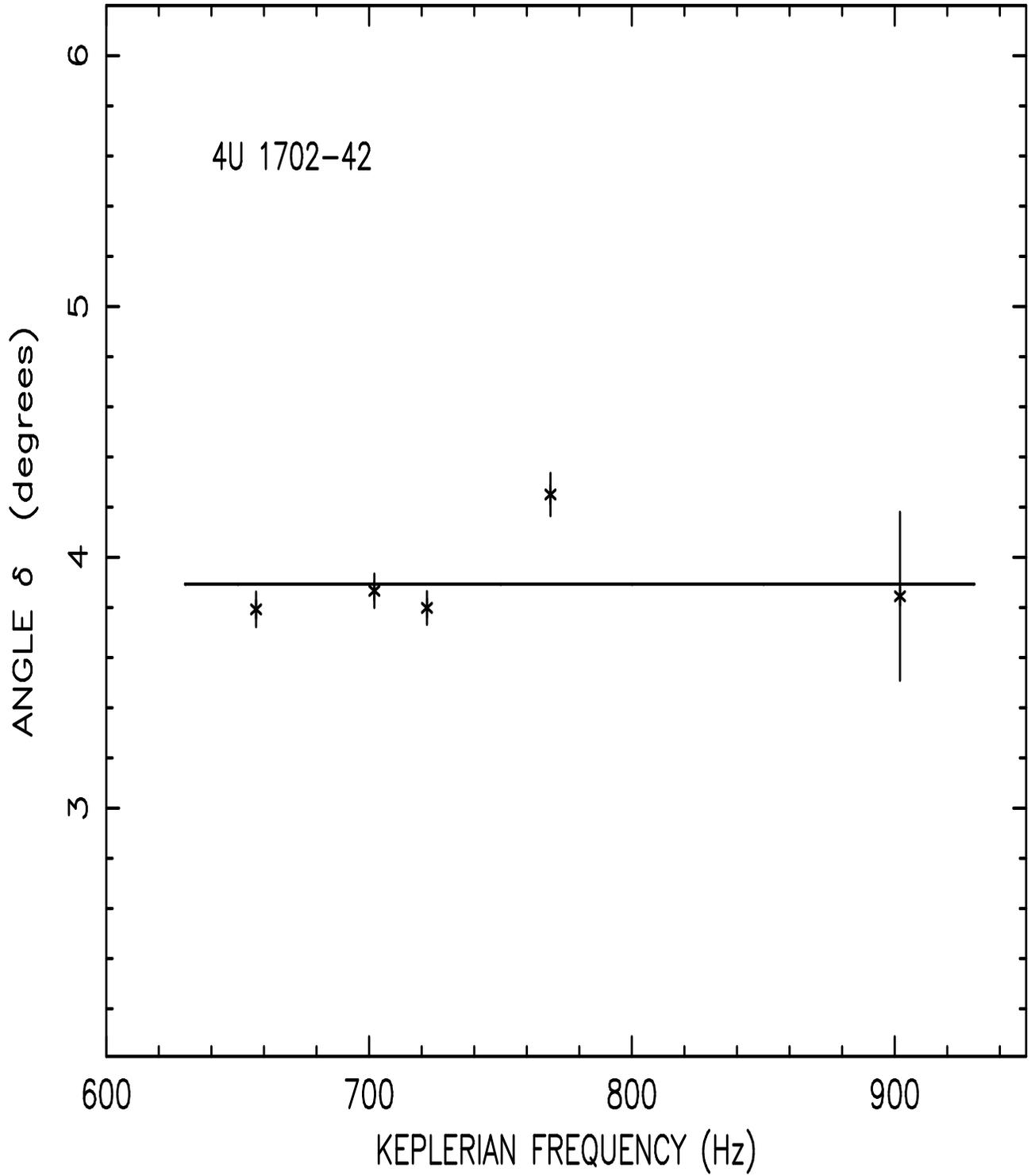
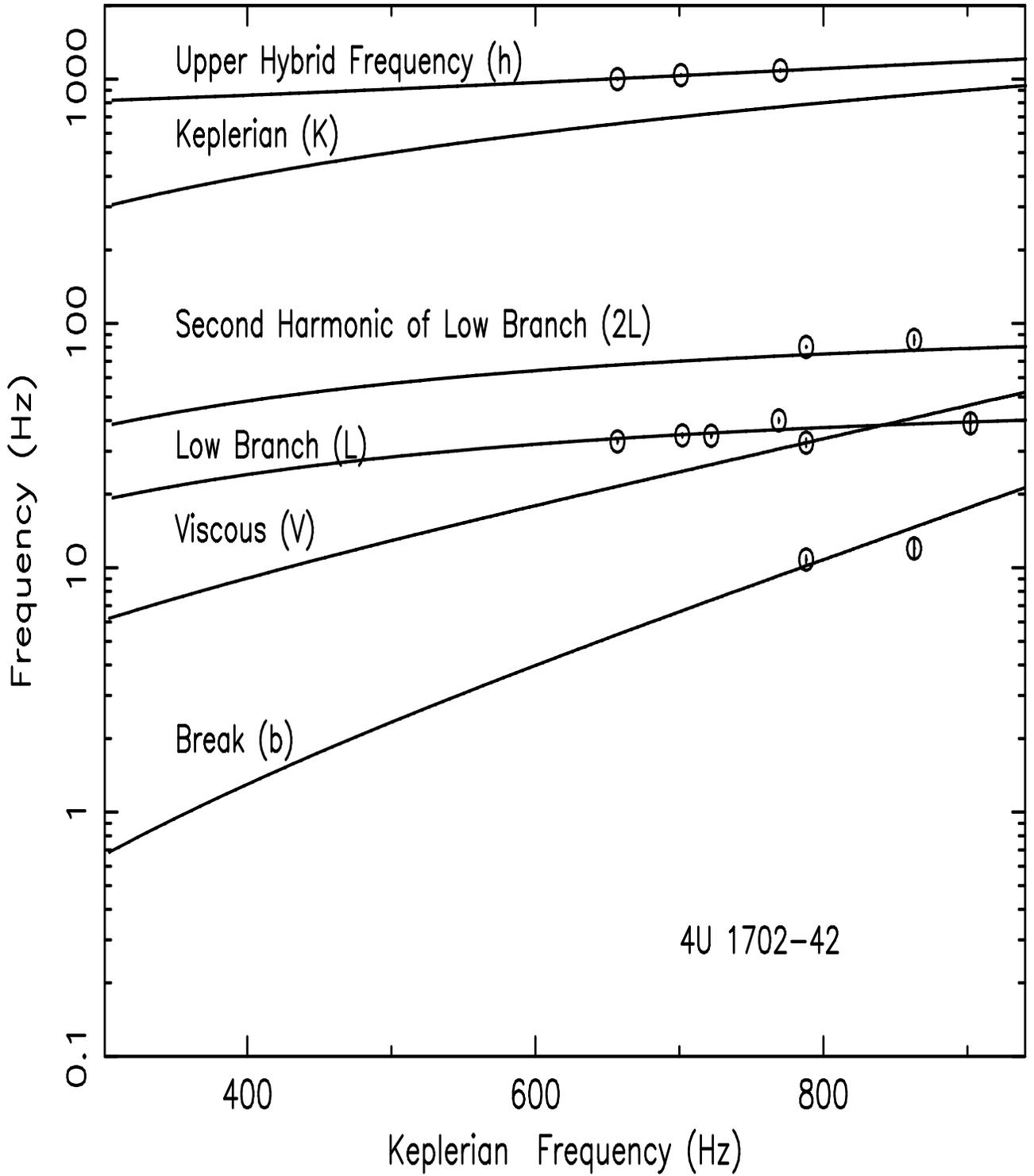
\end{document}